\documentclass[prd,12pt,nofootinbib,superscriptaddress]{revtex4-2}
\usepackage{bm}
\usepackage{amsmath}
\usepackage{amssymb}
\usepackage{graphicx}
\usepackage{float}
\usepackage{hyperref}
\usepackage{comment}
\usepackage{color}
\usepackage{orcidlink}
\usepackage{ulem}
\usepackage{subfigure}
\hypersetup{colorlinks=true,linkcolor=red,citecolor=blue}

\begin{document}
\title{On the evidence of dynamical dark energy}
\author{Qing Gao \orcidlink{0000-0003-3797-4370}}
\email{gaoqing1024@swu.edu.cn}
\affiliation{School of Physical Science and Technology, Southwest University,
Chongqing 400715, China}
\author{Zhiqian Peng}
\affiliation{School of Physical Science and Technology, Southwest University,
Chongqing 400715, China}
\author{Shengqing Gao\,\orcidlink{0000-0002-7501-2570}}
\email{gaoshengqing@hust.edu.cn}
\affiliation{School of Physics, Huazhong University of Science and Technology, 1037 LuoYu Rd, Wuhan, Hubei 430074, China}
\author{Yungui Gong \orcidlink{0000-0001-5065-2259}}
\email{Corresponding author. gongyungui@nbu.edu.cn}
\affiliation{Institute of Fundamental Physics and Quantum Technology, Department of Physics, School of Physical Science and Technology, Ningbo University, Ningbo, Zhejiang 315211, China}
\begin{abstract}
To elucidate the robustness of the baryon acoustic oscillation (BAO) data measured by the Dark Energy Spectroscopic Instrument (DESI) in capturing the dynamical behavior of dark energy,
we assess the model dependence of the evidence for dynamical dark energy inferred from the DESI BAO data.
While the DESI BAO data slightly tightens the constraints on model parameters and increases the tension between the Chevallier-Polarski-Linder (CPL) model and the $\Lambda$CDM model,
we find that the influence of DESI BAO data on the constraint of $w_0$
is small in the SSLCPL model.
In comparison to the CPL model, the tension with the $\Lambda$CDM model is reduced for the SSLCPL model, suggesting that the evidence for dynamical dark energy from DESI BAO data is dependent on cosmological models.
The inclusion of spatial curvature has little impact on the results in the SSLCPL model.
\end{abstract}

\maketitle

\section{Introduction}
Since the discovery of accelerated expansion of the Universe by observations of type Ia supernovae (SNe Ia) in 1998 \cite{SupernovaSearchTeam:1998fmf,SupernovaCosmologyProject:1998vns},
the nature of dark energy remains one of the most profound mysteries in modern cosmology.
Despite being one of the simplest candidates for dark energy, the expected value of the cosmological constant from vacuum energy is larger than the observed value by $10^{120}$ \cite{Weinberg:1988cp},
and the $\Lambda$CDM model faces the fine tuning and coincidence problems.
Additionally, the Hubble constant derived from observations of cosmic microwave background (CMB) anisotropies using the $\Lambda$CDM model \cite{Planck:2018vyg},
is $5.3\sigma$ away from the local measurements obtained through SNe Ia observations \cite{Riess:2022mme}.

Recent advancements in observations,
particularly through the Data Release 1 (DR1) from the first year of observations of  baryon acoustic oscillations (BAO) by the Dark Energy Spectroscopic Instrument (DESI),
have provided new insights into the property of dark energy,
specifically its dynamical aspects \cite{DESI:2024mwx}.
Using the flat Chevallier-Polarski-Linder (CPL) model \cite{Chevallier:2000qy,Linder:2002et},
DESI BAO data gives $w_0=-0.55^{+0.39}_{-0.21}$ at the $1\sigma$ confidence level and $w_a<-1.32$ at the 95\% confidence level \cite{DESI:2024mwx},
suggesting a mild tension with the $\Lambda$CDM model.
The combination of DESI BAO and Planck 2018 data gives $w_0=-0.45^{+0.34}_{-0.21}$
and $w_a=-1.79^{+0.48}_{-1.0}$,
indicating a preference for dynamical dark energy at the $\sim 2.6\sigma$ significance level \cite{DESI:2024mwx}.
Furthermore, the CPL model is favored over the $\Lambda$CDM model at the $2.5\sigma$, $3.5\sigma$ and $3.9\sigma$ significance level using the combination of DESI BAO, CMB \cite{Planck:2018vyg} and Pantheon Plus SNe Ia \cite{Scolnic:2021amr}, the combination of DESI BAO, CMB and Union3 SNe Ia \cite{Rubin:2023ovl}, and the combination of DESI BAO, CMB and SNe Ia data discovered and measured during the full five year of the Dark Energy Survey (DES) program \cite{Abbott:2024agi}, respectively \cite{DESI:2024mwx}.
For more discussions on the evidence for dynamical dark energy, please see Refs \cite{Cortes:2024lgw,Shlivko:2024llw,Giare:2024gpk,deCruzPerez:2024shj,Chan-GyungPark:2024spk,Roy:2024kni,Chatrchyan:2024xjj,Perivolaropoulos:2024yxv,Lu:2024hvv,Linder:2024rdj,Payeur:2024dnq} and references therein.

However, the evidence for evolving dark energy depends on the underlying cosmological models such as the $\Lambda$CDM model or the CPL model.
For a slowly rolling scalar field, the dynamics of thawing fields over a large redshift range can be approximated by the CPL parameterization $w(z)=w_0+w_a(1-a)$ with an explicit dengeneracy relation between $w_0$ and $w_a$, where $a$ is the scale factor \cite{Gao:2012ef,Gong:2013bn}.
The so called SSLCPL parameterization models a wide class of thawing scalar fields with only one free parameter $w_0$ \cite{Gao:2012ef,Gong:2013bn}.
In this paper, we explore the model dependence of the evidence for dynamical dark energy as inferred from DESI BAO data using the SSLCPL model.
We further investigate the impact of spatial curvature on the results.

This paper is organized as follows: Section \ref{sec2} discusses the observational data and cosmological models employed in our analysis,
while Section \ref{sec3} presents our results and their implications.
We conclude the paper in Section \ref{sec4}.

\section{Observational data and cosmology models}
\label{sec2}

\subsection{Observational data}\label{sec2.1}
For the BAO data, we use the BAO measurements in the redshift range of $0.1<z<4.2$, from the first year of observations with the DESI \cite{DESI:2024mwx}, and we label the dataset as BAO.
DESI is carrying out a five-year survey across 14,200 square degrees and uses six different galaxy tracers, including bright galaxies from the low-redshift survey, luminous red galaxies, emission line galaxies, quasars as direct tracers, and Lyman-$\alpha$ forest quasars,
to trace the distribution of neutral hydrogen.
DESI BAO measures the volume-averaged distance $D_V(z)$ in terms of the quantity $D_V(z)/r_d$ at two effective redshifts $z_\text{eff}=0.30$
and $z_\text{eff}=1.48$,
the tansverse comvoing angular diameter distance $D_M(z)$ in terms of $D_M(z)/r_d$ and
$D_H(z)/r_d$ at five different redshits, where $D_H(z)=1/H(z)$ and $r_d=r_s(z_d)$ is the sound horizon at the drag epoch $z_d$,
the drag redshift $z_d$ is fitted as \cite{Hu:1995en},
\begin{eqnarray}
\label{b1eq}
\label{zdfiteq} z_d&=&\frac{1291(\Omega_{m0}
h^2)^{0.251}}{1+0.659(\Omega_{m0} h^2)^{0.828}}\frac{1+b_1(\Omega_b h^2)^{b_2}}{0.962},\\
b_1&=&0.313(\Omega_{m0} h^2)^{-0.419}[1+0.607(\Omega_{m0}
h^2)^{0.674}], \\
b_2&=&0.238(\Omega_{m0} h^2)^{0.223},
\end{eqnarray}
the comoving sound horizon is
\begin{equation}
\label{rshordef}
r_s(z)=\int_z^\infty \frac{c_s(x)dx}{H(x)},
\end{equation}
the sound speed $c_s(z)=1/\sqrt{3[1+\bar{R_b}/(1+z)}]$,
$\bar{R_b}=3\Omega_b h^2/(4\times 2.469\times 10^{-5})$
and $h=H_0/(100\text{ km s}^{-1}\text{Mpc}^{-1})$.
Since the DESI BAO data points at the redshift $z=0.51$ show a statistical fluctuation  \cite{Colgain:2024xqj,DESI:2024mwx}, so we also use the DESI BAO data without the data points at the redshift $z=0.51$ to avoid the problem and label the dataset excluding the data points at the redshift $z=0.51$ as BAO$^{-}$.

For the CMB data from the Planck 2018 legacy release \cite{Planck:2018nkj,Planck:2018vyg}, instead of the full temperature anisotropy and polarization power spectra dataset,
we use the results of compressed parameters given in the Table F1 in Ref. \cite{Rubin:2023ovl},
the sift parameter $R$ at recombination with the redshift $z_{*}$,
\begin{equation}
\label{shiftdef}
R=\sqrt{\Omega_{m0}H^2_0}\,D_M(z_*),
\end{equation}
the acoustic angular scale $\theta_*=r_*/D_M(z_*)$ at recombination,
and the baryon density $\omega_b=\Omega_b h^2$, where $r_*=r_s(z_*)$ is the sound horizon at recombination,
and the recombination redshift $z^{*}$ is fitted as \cite{Hu:1995en},
\begin{eqnarray}
\label{zstareq}
z^{*}&=&1048[1+0.00124(\Omega_b
h^2)^{-0.738}][1+g_1(\Omega_{m0} h^2)^{g_2}],\\
g_1&=&\frac{0.0783(\Omega_b h^2)^{-0.238}}{1+39.5(\Omega_b
h^2)^{0.763}},\\
g_2&=&\frac{0.560}{1+21.1(\Omega_b h^2)^{1.81}}.
\end{eqnarray}
We label the compressed Planck data as P18.

We use three different SNe Ia data: 1829 SNe Ia compiled by DES \cite{DES:2024jxu}, the Union3 compilation of 2087 SNe Ia \cite{Rubin:2023ovl}, and the Pantheon Plus sample of 1550 spectroscopically confirmed SNe Ia \cite{Scolnic:2021amr}.
The DES SNe Ia dataset, labeled as D5, includes 1635 photometrically-classified DES SNe Ia in the redshift range $0.10<z<1.13$ and 194 low-redshift SNe Ia with redshifts $0.025<z<0.1$.
The Union3 SNe Ia dataset, labeled as U3, comes from 24 datasets spanning the redshift range $0.01<z<2.26$.
The Pantheon Plus SNe Ia dataset, labeled as PP, covers the redshift range $0.001<z<2.26$.
In order to mitigate the effects of peculiar velocity corrections, a bound $z>0.01$ was imposed, resulting in a subset of 1590 SNe Ia data points with redshifts $0.01016<z<2.26$ in the Pantheon Plus SNe Ia dataset.

The Hubble parameter $H(z)$ data compiled in Ref. \cite{Gadbail:2024rpp}, labeled as H, includes 32 $H(z)$ data points obtained with the cosmic chronometer
 (CCH) method using the differential redshift time derived from the
the age evolution of passively evolving galaxies \cite{Jimenez:2001gg,Simon:2004tf,Stern:2009ep,Zhang:2012mp,Moresco:2012jh,Moresco:2015cya,Moresco:2016mzx,Ratsimbazafy:2017vga,Borghi:2021rft},
and 26 data points obtained from radial BAO observations, covering the redshits $0.07<z<2.36$ \cite{Gaztanaga:2008xz,Chuang:2012qt,Blake:2012pj,BOSS:2012gof,BOSS:2013rlg,Oka:2013cba,BOSS:2013igd,BOSS:2014hwf,BOSS:2016zkm,BOSS:2016wmc,BOSS:2017fdr}.

\subsection{Methods}
\label{sec2.2}
We perform Markov Chain Monte Carlo (MCMC) simulations by using the publicly available emcee Python code \cite{Foreman-Mackey:2012any} and analyze the samples using the GetDist Python module \cite{Lewis:2019xzd} to give constraints on model parameters.
We combine the CMB and $H(z)$ data with different SNe Ia data to get the datasets: P18+H+PP, P18+H+D5 and P18+H+U3,
to derive the average values, confidence levels and likelihood distributions of model parameters for various models.
Then we combine the DESI BAO data including and excluding the data points at the redshfit $z=0.51$ with the above datasets to obtain datasets: BAO+P18+H+PP, BAO+P18+H+D5, BAO+P18+H+U3, BAO$^-$+P18+H+PP, BAO$^-$+P18+H+D5 and BAO$^-$+P18+H+U3, and use the combined data to constrain the model parameters of various models.

To assess the performance of the models, we calculate
the Akaike information criterion (AIC),
\begin{equation}
\text{AIC}=\chi^2_{min}+2m,
\end{equation}
where $\chi^2_{min}$ is the minimum value of $\chi^2$ for the best-fit cosmological parameters and $m$ is the number of independent cosmological parameters.
We compare the performance of CPL and SSLCPL models with the $\Lambda$CDM model
by calculating the difference in the value of AIC between the CPL and SSLCPL model and that of $\Lambda$CDM model.
If $-2\leq\Delta\text{AIC}<0$, then the model is favored over the $\Lambda$CDM model with weak evidence.
If $-6\leq\Delta$AIC, then the evidence in favor of the model is positive.
If $-10\leq\Delta$AIC, then there is strong evidence in favor of the model.
When $\Delta \text{AIC}<-10$, there is very
strong evidence in favor of the model against the $\Lambda$CDM model \cite{deCruzPerez:2024shj}.

\subsection{Cosmological models}\label{sec2.3}
The luminosity distance is
\begin{equation}
\label{dleq}
d_L=\frac{1+z}{H_0\sqrt{|\Omega_{k}|}}\sin\left[\sqrt{|\Omega_{k}|}\int_0^z \frac{dx}{E(x)}\right],
\end{equation}
where $E(z)=H(z)/H_0$ and
\begin{equation}
\frac{{\rm sinn}(\sqrt{|\Omega_k|}x)}{\sqrt{|\Omega_k|}}=\begin{cases}
\sin(\sqrt{|\Omega_k|}x)/\sqrt{|\Omega_k|},& {\rm if}\ \Omega_k<0,\\
x, & {\rm if}\  \Omega_k=0, \\
\sinh(\sqrt{|\Omega_k|}x)/\sqrt{|\Omega_k|}, & {\rm if}\  \Omega_k>0.
\end{cases}
\end{equation}
The transverse comoving distance $D_M(z)=d_L(z)/(1+z)$.
The volume-averaged distance $D_V(z)=[zD_M(z)^2 D_H(z)]^{1/3}$.

For the $\Lambda$CDM model,
\begin{equation}
\label{lcdmez1}
E^2(z)=\Omega_{r}(1+z)^4+\Omega_{m0}(1+z)^3+\Omega_k (1+z)^2+1-\Omega_{m0}-\Omega_{r}-\Omega_k.
\end{equation}

For Chevallier-Polarski-Linder (CPL) parametrization,
\begin{equation}
\label{cpleq1}
w(z)=w_0+\frac{w_a z}{1+z},
\end{equation}
where $w_0$ and $w_a$ are model parameters,
\begin{equation}
\label{cplez1}
\begin{split}
E^2(z)=&\Omega_{r}(1+z)^4+\Omega_{m0}(1+z)^3+\Omega_k (1+z)^2\\
&+(1-\Omega_{m0}-\Omega_{r}-\Omega_k)(1+z)^{3(1+w_0+w_a)}\exp[-3w_a
z/(1+z)].
\end{split}
\end{equation}

The SSLCPL parametrization approximates the dynamics of general thawing scalar fields over a large redshift range with only one free parameter $w_0$, and reduces to $\Lambda$CDM model when the parameter $w_0=-1$ \cite{Gao:2012ef,Gong:2013bn}.
The SSLCPL parametrization is $w(a)=w_0+w_a(1-a)$ with
\begin{eqnarray}
\label{waeq1}
w_a=6(1+w_0)\frac{(\Omega_{\phi 0}^{-1}-1)[\sqrt{\Omega_{\phi0}}-\tanh^{-1}(\sqrt{\Omega_{\phi0}})]}
{\Omega_{\phi 0}^{-1/2}-(\Omega_{\phi 0}^{-1}-1)\tanh^{-1}(\sqrt{\Omega_{\phi0}})},
\end{eqnarray}
where $\Omega_{\phi 0}=1-\Omega_{m0}-\Omega_r-\Omega_k$.

\section{Results and Discussion}\label{sec3}
To verify if the compressed CMB data yields results consistent with the full power spectra data,
we compare the constraints on the $\Lambda$CDM and CPL models obtained from the combined CMB and BAO data, labeled as BAO+P18, and the results are shown in Table \ref{test}.
From Table \ref{test}, we see that the results are similar and consistent for $\Lambda$CDM and flat CPL models.
For the CPL model with spatial curvature, although the best fit value for $w_a$ using the full power spectra is smaller--especially for the constraint from the combination of DESI BAO and Planck 2018 data--
the results remain consistent with those obtained using the compressed data points at the
$1\sigma$ level.
These results confirm that we can use the three compressed data points to represent the full power spectra.

\begin{figure}[htp]
\centering
\includegraphics[width=0.9\textwidth]{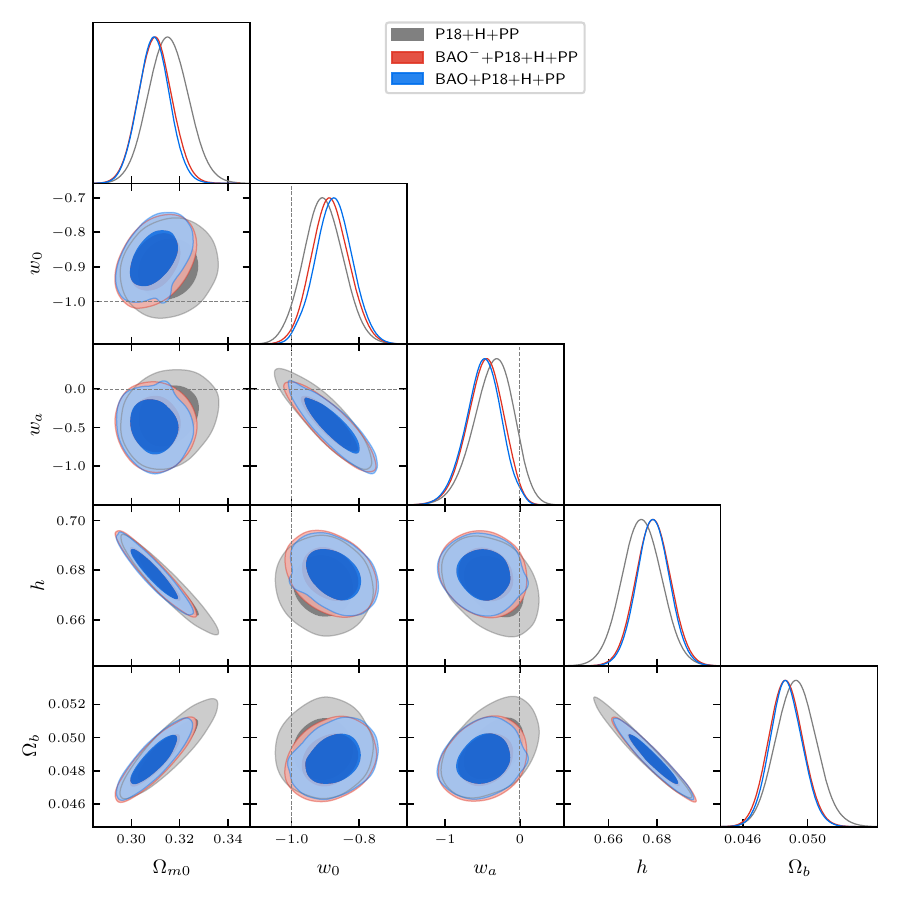}
\caption{Marginalized 68\% and 95\% posteriors on $\Omega_{m0}$, $w_0$, $w_a$ for the flat CPL model, from the combined P18+H+PP, BAO$^-$+P18+H+PP and BAO+P18+H+PP data in gray, red and blue, respectively.}
\label{cplpp}
\end{figure}

\begin{figure*}
\centerline{$\begin{array}{cc}
\includegraphics[width=0.45\linewidth]{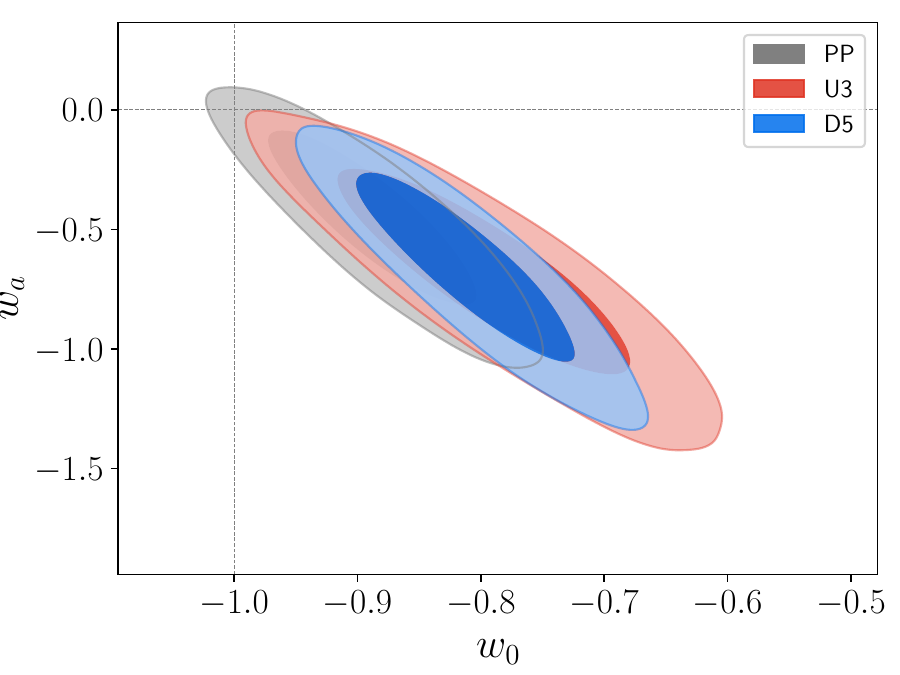}\quad \quad &
\includegraphics[width=0.45\linewidth]{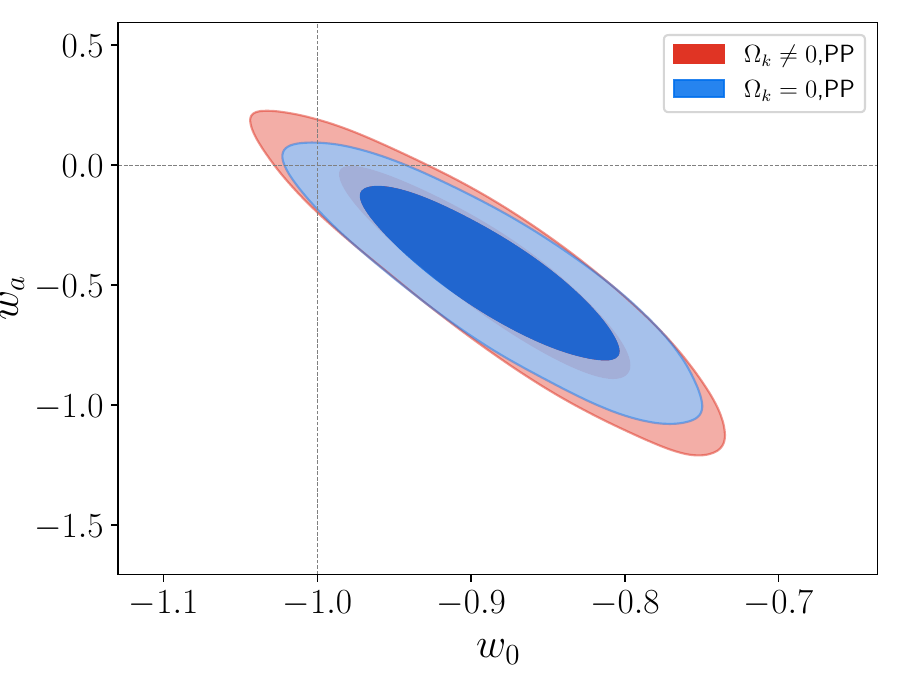}\\
\includegraphics[width=0.45\linewidth]{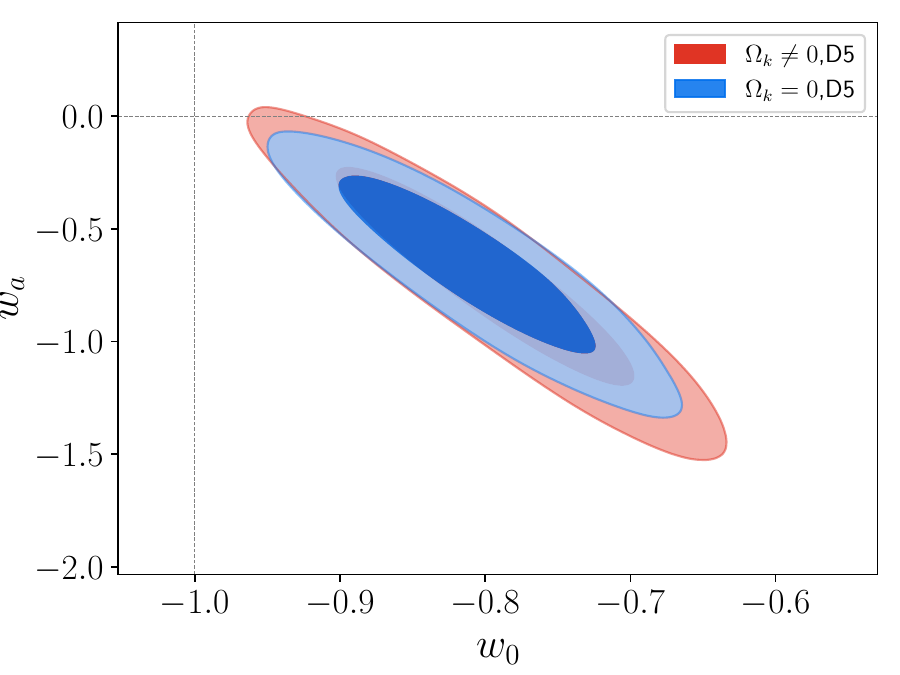} \quad \quad &
\includegraphics[width=0.45\linewidth]{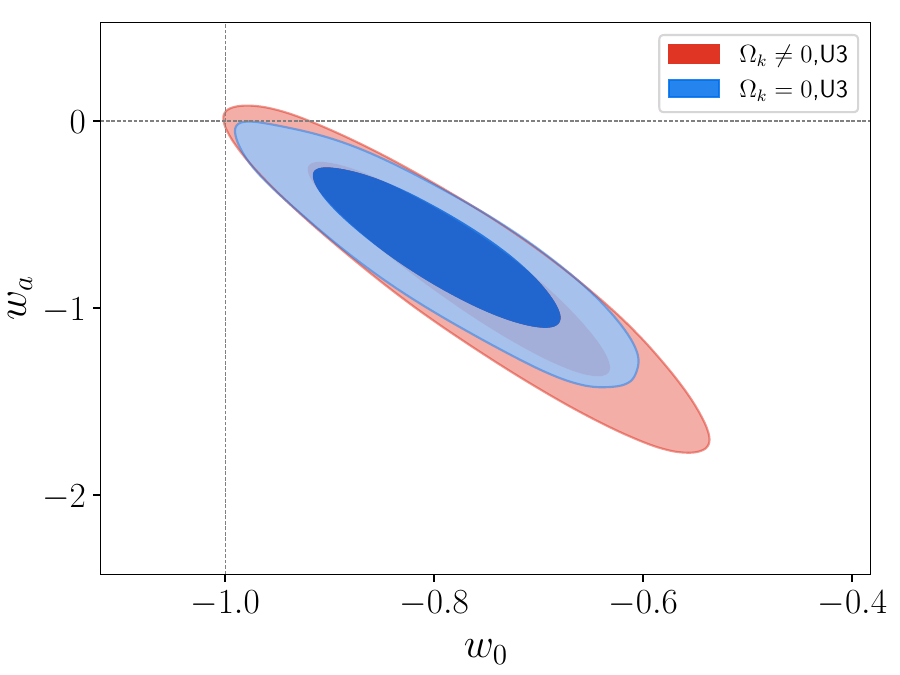}
\end{array}$}
\caption{
68\% and 95\% confidence contours of $w_0-w_a$ for the CPL model, from the combined BAO$^{-}$+P18+H and various SNe Ia data.
The upper left panel compares the results for the flat CPL model from the combinations with various SNe Ia data,
the upper right panel compares the results for flat and non-flat CPL models with the dataset BAO$^{-}$+P18+H+PP,
the lower left panel compares the results for flat and non-flat CPL models with the dataset BAO$^{-}$+P18+H+D5,
and the lower right panel compares the results for flat and non-flat CPL models with the dataset BAO$^{-}$+P18+H+U3.
}
\label{cpla}
\end{figure*}

\begin{figure}[htp]
\centerline{\includegraphics[width=0.9\textwidth]{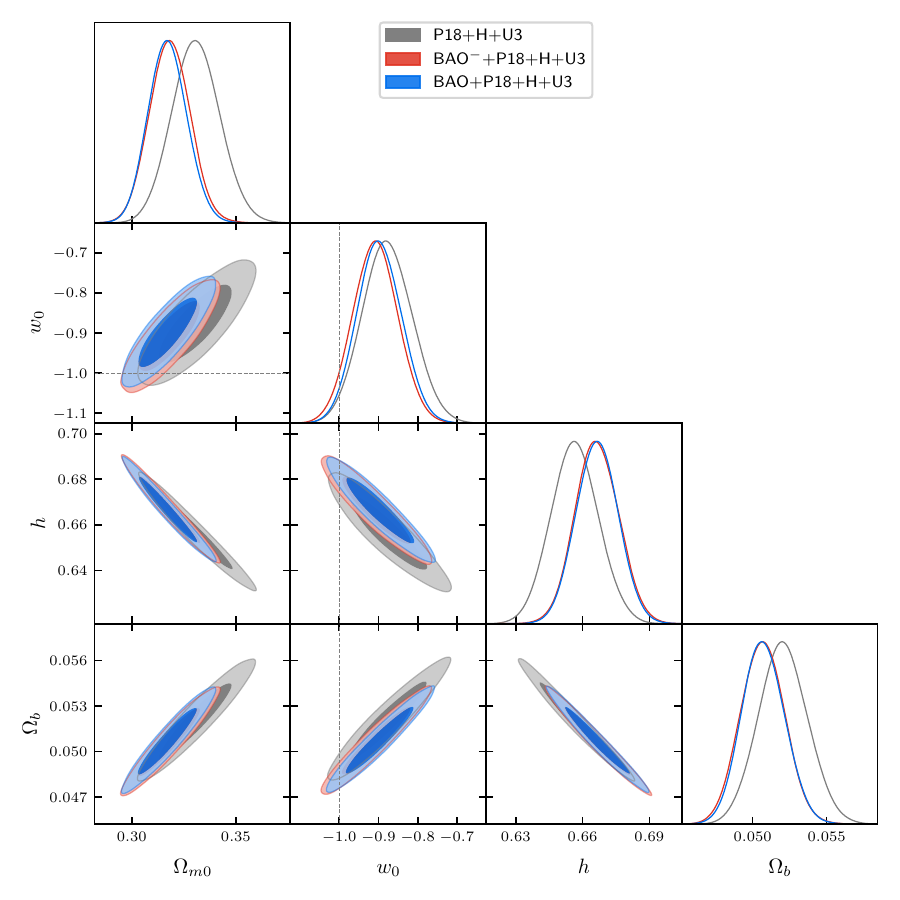}}
\caption{Marginalized 68\% and 95\% posteriors on $\Omega_{m0}$, $w_0$ for the flat SSLCPL model, from the combined P18+H+U3, BAO$^-$+P18+H+U3 and BAO+P18+H+U3 data in gray, red and blue, respectively.
}\label{sslu3}
\end{figure}

\begin{figure}[htp]
\centerline{\includegraphics[width=0.9\textwidth]{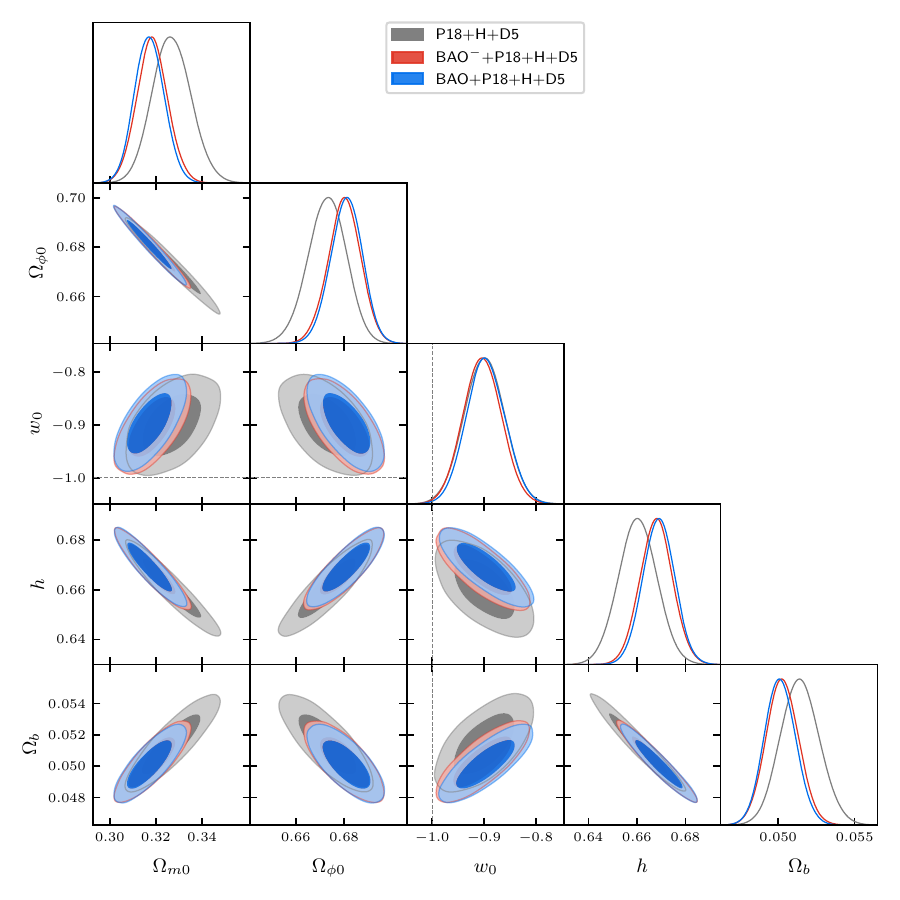}}
\caption{Marginalized 68\% and 95\% posteriors on $\Omega_{m0}$,$\Omega_{\phi0}$, $w_0$ for non-flat SSLCPL model, from the combined P18+H+D5, BAO$^-$+P18+H+D5 and BAO+P18+H+D5 data in gray, red and blue, respectively.
}\label{sslkd5}
\end{figure}

\begin{figure}
\centerline{$\begin{array}{cc}
\includegraphics[width=0.45\textwidth]{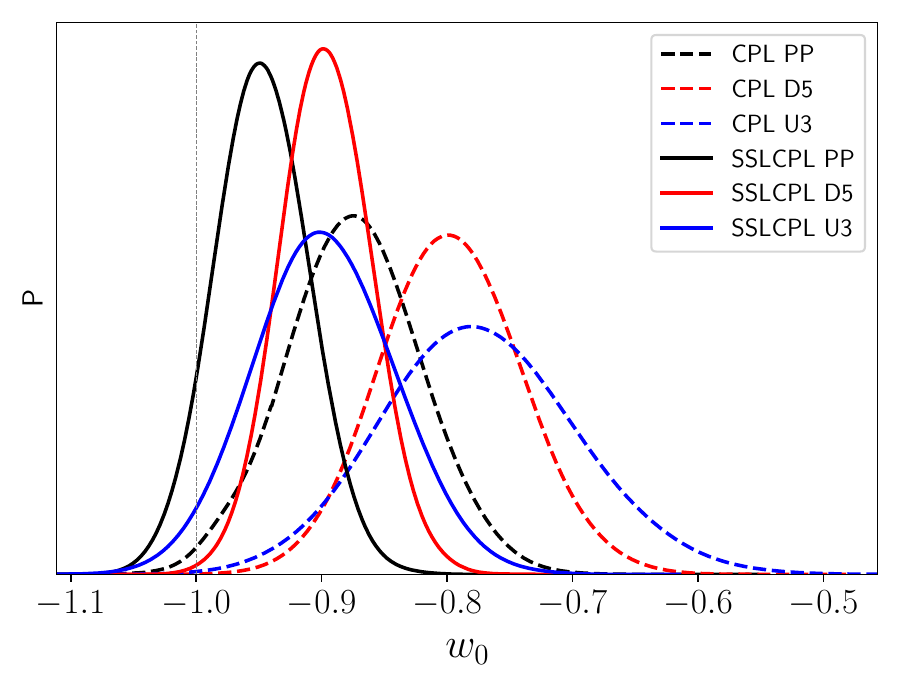}&
\includegraphics[width=0.45\textwidth]{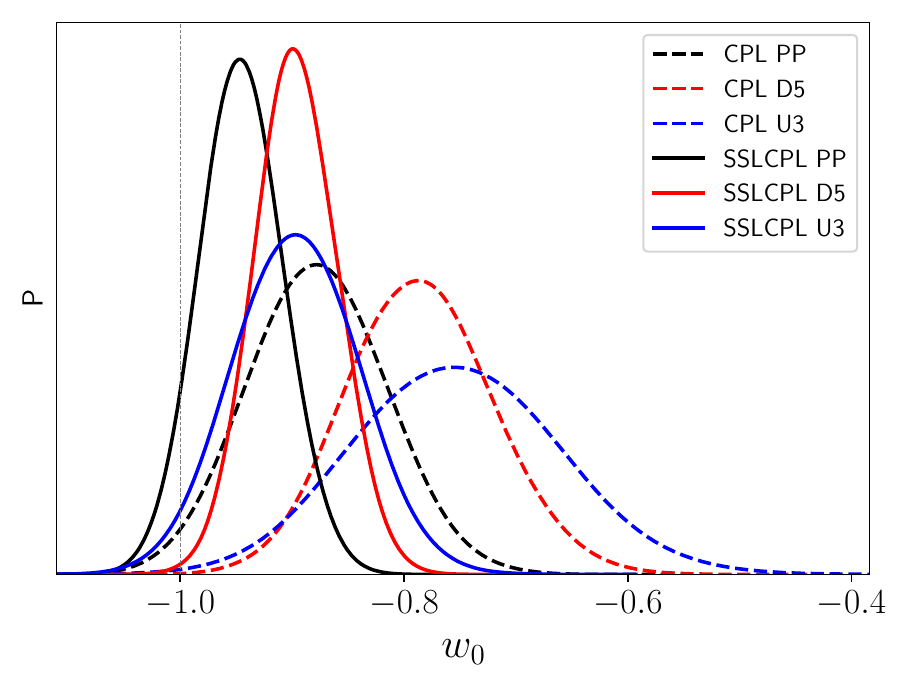}
\end{array}$}
\caption{The marginalized 1D posteriors on $w_0$ for the CPL and SSLCPL models, from the combined BAO+P18+H+PP, BAO+P18+H+D5 and BAO+P18+H+U3 data.
The left panel shows the results for the flat CPL and SSLCPL models, and
the right panel shows the results for the non-flat CPL and SSLCPL models.
The dashed and solid lines are for CPL and SSLCPL models, respectively.
The black, red and blue lines are for the combination BAO+P18+H+PP, BAO+P18+H+D5 and BAO+P18+H+U3, respectively.}
\label{pw0}
\end{figure}

We fit the flat CPL model to the combined data: P18+H+PP, P18+H+D5 and P18+H+U3,
then we add BAO and BAO$^{-}$ data to the combined dataset to assess the impact of BAO data on the constraints. We repeat the process with non-flat CPL model to see the effect of the spatial curvature on the results.
Finally, we replace the CPL model with the SSLCPL model to evaluate whether the conclusion depends on cosmological models.
We fit both flat and non-flat CPL and SSLCPL models to  nine combinations of data: P18+H+PP, P18+H+D5, P18+H+U3, BAO+P18+H+PP, BAO+P18+H+D5, BAO+P18+H+U3, BAO$^{-}$+P18+H+PP, BAO$^{-}$+P18+H+D5 and BAO$^{-}$+P18+H+U3. The results are shown in Table \ref{cpl} and \ref{ssl}, and Figs. \ref{cplpp}-\ref{pw0}.

As shown in Table \ref{cpl}, Figs. \ref{cplpp} and \ref{cpla}, for the flat CPL model,
the best fit values of $\Omega_{m0}$ and $w_0$ are smallest with the combined P18+H+PP data, while they are largest with the combined P18+H+U3 data.
The best fit value of $w_a$ is largest with the combined P18+H+PP data, while it is smallest or further away from $0$ with the combined P18+H+U3 data.
The error bars with the combined P18+H+U3 data are a little larger than those obtained with the other two combinations.
The best fit value for $w_0$ and $w_a$ are both negative.
The results with the combined P18+H+PP data are more consistent with the $\Lambda$CDM model compared with those obtained from the combinations with D5 and U3.
Adding the DESI BAO data, with or without the data points at the redshift $z=0.51$, to the combined CMB, $H(z)$ and SNe Ia data,
we observe the same trend for the best fit values of the model parameters obtained from different SNe Ia data,
with the constraints on these parameters becoming slightly more stringent.
With the data points at the redshift $z=0.51$, the addition of BAO data decreases the values of $\Omega_{m0}$ and $w_a$, and increases $w_0$, thereby intensifying the tension with the $\Lambda$CDM model.
In the absence of the data points at the redshift $z=0.51$,
the results are similar, although the tension with $\Lambda$CDM diminishes slightly, indicating that the impact of data points at the redshift $z=0.51$ is small.
By introducing spatial curvature as a free parameter, the constraints on the parameters $\Omega_{m0}$, $w_0$ and $w_a$ broaden a little compared to those in the flat case and the same trend for the constraints with different SNe Ia data continues to persist.
With the combined BAO+P18+H+D5 data,
we get $w_0=-0.797\pm 0.057$ and $w_a=-0.70^{+0.27}_{-0.23}$ for the flat CPL model,
$w_0=-0.786\pm 0.066$ and $w_a=-0.76^{+0.34}_{-0.28}$ for the non-flat flat CPL model,
the values of $\Delta$AIC reach $-10$ and $-8$ for the flat and non-flat cases, respectively,
indicating strong evidence for dynamical dark energy.
The constraints on $w_0$ and $w_a$ from the combined BAO+P18+H+U3 data are similar,
although the values of $\Delta$AIC are smaller.

As shown in Table \ref{ssl} and Figs. \ref{sslu3}-\ref{pw0}, for the SSLCPL model,
the constraints on $w_0$ from the combination with D5 and U3 are similar, although the error bars on $w_0$ from the combination with U3 are bigger that those from the combination with D5.
The constraints on $w_0$ obtained from the combined data with PP are consistent with the $\Lambda$CDM model at around the $1\sigma$ level,
while the constraints on $w_0$ from the combined data with D5 and U3 are consistent with the $\Lambda$CDM model at around the $2\sigma$ level.
Without the DESI BAO data points at the redshift $z=0.51$ as shown in Figs. \ref{sslu3} and \ref{sslkd5}, the value of $w_0$ moves closer to $-1$.
The impact of the spatial curvature on the constraints of $w_0$ is negligible.
In summary, compared with the CPL model as shown in Fig. \ref{pw0},
the tension with the $\Lambda$CDM model is reduced a little bit for the SSLCPL model.
Even with the combined BAO+P18+H+D5 data, $w_0=-0.897\pm 0.037$ and $\Delta$AIC is $-6$ for both flat and non-flat SSLCPL models.

\begin{table*}[htbp]
	\renewcommand\tabcolsep{4.0pt}
    \centering
	\begin{tabular}{lcccc}
		\hline
		\hline
		Model/Data   &$\Omega_{m0}$ &$\Omega_\Lambda$& $w_0$ & $w_a$\\

		\hline
        \bf{Flat $\bm{\Lambda}$CDM}  &\\
 	BAO+P18   ~& $0.3023\pm0.0054$~& $-$~& $-$~&$-$\\
    {\it DESI}    ~& $0.3069\pm0.0050$~& $-$~& $-$~&$-$\\
        \hline
        $\bm{\Lambda}$\bf{CDM}+$\bm{\Omega_k}$  &\\
 	BAO+P18   ~& $0.2999\pm0.0056$~& $0.6968\pm0.0055$~& $-$~&$-$\\
    {\it DESI}   ~& $0.3049\pm0.0051$~& $0.6927\pm0.0053$~& $-$~&$-$\\
        \hline
        \bf{Flat CPL}  &\\
        BAO+P18   ~& $0.334^{+0.031}_{-0.019}$~& $-$~& $-0.55^{+0.32}_{-0.12}$~&$-1.52^{+0.51}_{-0.87}$\\
        {\it DESI}   ~& $0.344^{+0.032}_{-0.027}$~& $-$~& $-0.45^{+0.34}_{-0.21}$~&$-1.79^{+0.48}_{-1.0}$\\

        BAO+P18+PP   ~& $0.3082\pm0.0068$~& $-$~& $-0.837\pm0.065$~&$-0.67^{+0.32}_{-0.27}$\\
        {\it DESI}   ~& $0.3085\pm0.0068$~& $-$~& $-0.827\pm0.063$~&$-0.75^{+0.29}_{-0.25}$\\

        BAO+P18+U3   ~& $0.3233\pm0.0097$~& $-$~& $-0.65\pm0.11$~&$-1.23^{+0.44}_{-0.38}$\\
        {\it DESI}   ~& $0.3230\pm0.0095$~& $-$~& $-0.65\pm0.10$~&$-1.27^{+0.40}_{-0.34}$\\

        BAO+P18+D5   ~& $0.3162\pm0.0066$~& $-$~& $-0.735\pm0.07$~&$-0.99^{+0.34}_{-0.30}$\\
        {\it DESI}   ~& $0.3160\pm0.0065$~& $-$~& $-0.727\pm0.067$~&$-1.05^{+0.31}_{-0.27}$\\
        \hline
        \bf{CPL+$\bm{\Omega_k}$}  &\\
        BAO+P18   ~& $0.331^{+0.033}_{-0.021}$~& $0.667^{+0.021}_{-0.032}$~& $-0.61^{+0.37}_{-0.15}$~&$-1.24^{+0.62}_{-1.0}$\\
        {\it DESI}   ~& $0.347^{+0.031}_{-0.025}$~& $0.6539^{+0.031}_{-0.025}$~& $-0.41^{+0.33}_{-0.18}$~&$<-1.61$\\

        BAO+P18+PP   ~& $0.3083\pm0.0068$~& $0.6893\pm0.0071$~& $-0.875\pm0.07$~&$-0.44^{+0.36}_{-0.30}$\\
        {\it DESI}   ~& $0.3084\pm0.0067$~& $0.6913\pm0.0069$~& $-0.831\pm0.066$~&$-0.73^{+0.32}_{-0.28}$\\

        BAO+P18+U3   ~& $0.3225\pm0.0098$~& $0.6758\pm0.0097$~& $-0.69\pm0.11$~&$-1.01^{+0.49}_{-0.44}$\\
        {\it DESI}   ~& $0.3233^{+0.0089}_{-0.010}$~& $0.6771\pm0.0091$~& $-0.64\pm 0.11$~&$-1.30^{+0.45}_{-0.39}$\\

        BAO+P18+D5   ~& $0.3159\pm0.0066$~& $0.6823\pm0.0068$~& $-0.766\pm0.077$~&$-0.80^{+0.41}_{-0.35}$\\
        {\it DESI}   ~& $0.3163\pm0.0065$~& $0.6839\pm0.0068$~& $-0.725\pm0.071$~&$-1.06^{+0.35}_{-0.31}$\\
		\hline
		\hline
	\end{tabular}
	\caption{Comparison of the results with the compressed Planck data and the full power spectra. The results labeled as {\it DESI} are copied from Ref. \cite{DESI:2024mwx}.}
\label{test}
\end{table*}

\begin{table*}[htbp]
	\renewcommand\tabcolsep{4.0pt}
    \centering
	\begin{tabular}{lccccc}
		\hline
		\hline
		Data & $\Delta$AIC  &$\Omega_{m0}$ &$\Omega_{\phi 0}$& $w_0$ & $w_a$\\

		\hline

        P18+H+PP  &$2$ ~& $0.3154\pm0.0082$~& $-$ & $-0.905\pm0.059$~&$-0.36^{+0.29}_{-0.24}$\\
          &$2$ ~& $0.3173\pm0.0087$~& $0.6846\pm0.0085$~& $-0.869^{+0.066}_{-0.076}$~&$-0.58^{+0.42}_{-0.34}$\\

        P18+H+D5  &$-4$ ~& $0.3236\pm0.0079$~& $-$~&$-0.831\pm0.063$~&$-0.55^{+0.32}_{-0.26}$\\
          &$-6$ ~& $0.3269\pm0.0083$~& $0.6764\pm0.0079$~& $-0.763\pm0.082$~&$-0.96\pm0.44$\\

        P18+H+U3  &$-2$ ~& $0.328\pm0.011$~&$-$ ~&$-0.806\pm0.08$~&$-0.58^{+0.32}_{-0.27}$\\
          &$-4$ ~& $0.335\pm0.012$~& $0.668\pm0.012$~& $-0.69\pm0.11$~&$-1.16^{+0.57}_{-0.48}$\\

        BAO+P18+H+PP  &$-2$ ~& $0.3094\pm0.0065$~& $-$~&$-0.875\pm0.054$~&$-0.49^{+0.25}_{-0.22}$\\
          &$1$ ~& $0.3098\pm0.0065$~& $0.6899\pm0.0067$~& $-0.878\pm0.062$~&$-0.47^{+0.30}_{-0.27}$\\

        BAO+P18+H+D5  &$-10$ ~& $0.3177\pm0.0064$~& $-$~&$-0.797\pm0.057$~&$-0.70^{+0.27}_{-0.23}$\\
          &$-8$ ~& $0.3177\pm0.0065$~& $0.6827\pm0.0065$~& $-0.786\pm0.066$~&$-0.76^{+0.34}_{-0.28}$\\

         BAO+P18+H+U3  &$-4$ ~& $0.3208\pm0.0092$~&$-$ ~&$-0.775\pm0.078$~&$-0.74^{+0.31}_{-0.25}$\\
           &$-2$ ~& $0.3218\pm0.0095$~& $0.6788\pm0.0093$~& $-0.754\pm0.092$~&$-0.83^{+0.39}_{-0.34}$\\

        BAO$^-$+P18+H+PP  &$0$ ~& $0.3099\pm0.0069$~&$-$ ~&$-0.885\pm0.055$~&$-0.47^{+0.26}_{-0.22}$\\
          &$1$ ~& $0.3099\pm0.0069$~& $0.6902\pm0.007$~& $-0.886\pm0.063$~&$-0.46^{+0.31}_{-0.27}$\\

        BAO$^-$+P18+H+D5  &$-7$ ~& $0.318\pm0.0067$~& $-$~&$-0.809\pm0.058$~&$-0.67^{+0.28}_{-0.23}$\\
          &$-6$ ~& $0.3186\pm0.0068$~& $0.6819\pm0.0068$~& $-0.797\pm0.068$~&$-0.72^{+0.34}_{-0.30}$\\

        BAO$^-$+P18+H+U3  &$-3$ ~& $0.3199\pm0.0094$~&$-$ ~&$-0.792\pm0.08$~&$-0.70^{+0.32}_{-0.26}$\\
          &$-2$ ~& $0.3213\pm0.0098$~& $0.6795\pm0.0096$~& $-0.77\pm0.095$~&$-0.80^{+0.40}_{-0.34}$\\
		\hline
		\hline
	\end{tabular}
	\caption{The constraints on CPL model.}
\label{cpl}
\end{table*}

\begin{table*}[htbp]
	\renewcommand\tabcolsep{4.0pt}
    \centering
	\begin{tabular}{lcccc}
		\hline
		\hline
		Data & $\Delta$AIC  &$\Omega_{m0}$ &$\Omega_{\phi0}$& $w_0$ \\

		\hline

 	P18+H+PP  &$1$ ~& $0.3184\pm0.008$~&$-$ ~&$-0.951\pm0.04$\\
         &$1$ ~& $0.3186\pm0.0084$~& $0.6812\pm0.008$~& $-0.95\pm0.04$\\

        P18+H+D5  &$-4$ ~& $0.3268\pm0.0077$~&$-$ ~&$-0.9\pm0.039$\\
          &$-4$ ~& $0.3269\pm0.0084$~& $0.6729\pm0.0079$~& $-0.9\pm0.039$\\

        P18+H+U3  &$-2$ ~& $0.331\pm0.011$~&$-$ ~&$-0.877\pm0.064$\\
          &$-2$ ~& $0.332\pm0.012$~& $0.668\pm0.012$~& $-0.874\pm0.062$\\

        BAO+P18+H+PP  &$0$ ~& $0.3099\pm0.0064$~&$-$ ~&$-0.948\pm0.038$\\
          &$0$ ~& $0.3098\pm0.0066$~& $0.6885\pm0.0067$~& $-0.945\pm0.038$\\

        BAO+P18+H+D5  &$-6$ ~& $0.3174\pm0.0064$~&$-$ ~&$-0.897\pm0.037$\\
          &$-6$ ~& $0.3172\pm0.0064$~& $0.681\pm0.0065$~& $-0.897\pm0.037$\\

        BAO+P18+H+U3  &$-1$ ~& $0.3173\pm0.0092$~&$-$~&$-0.898\pm0.056$\\
          &$-1$ ~& $0.3172\pm0.0092$~& $0.6811\pm0.0093$~& $-0.896\pm0.056$\\

        BAO$^-$+P18+H+PP  &$1$ ~& $0.3113\pm0.0068$~&$-$ ~&$-0.954\pm0.038$\\
          &$1$ ~& $0.3108\pm0.0069$~& $0.6878\pm0.0069$~& $-0.953\pm0.038$\\

        BAO$^-$+P18+H+D5  &$-5$ ~& $0.3192\pm0.0067$~&$-$ ~&$-0.903\pm0.037$\\
          &$-5$ ~& $0.3184\pm0.0067$~& $0.6801\pm0.0067$~& $-0.903\pm0.037$\\

        BAO$^-$+P18+H+U3  &$-1$ ~& $0.3184\pm0.0096$~& $-$~&$-0.908\pm0.058$\\
          &$-1$ ~& $0.3177\pm0.0097$~& $0.6809\pm0.0097$~& $-0.908\pm0.058$\\
		\hline
		\hline
	\end{tabular}
	\caption{The constraints on SSLCPL model.}
\label{ssl}
\end{table*}

\section{Conclusion}
\label{sec4}
For the flat CPL model, the best fit values of $\Omega_{m0}$ and $w_0$ are smallest with the combined P18+H+PP data, while they are largest with the combined P18+H+U3 data.
The best fit value of $w_a$ is largest with the combined P18+H+PP data, while it is smallest or further away from $0$ with the combined P18+H+U3 data.
The error bars on $w_0$ from the combination with U3 are a little larger than those obtained with the other two combinations.
Adding the DESI BAO data, with or without the data points at the redshift $z=0.51$, to the combined CMB, $H(z)$ and SNe Ia data,
the constraints on model parameters become slightly more stringent.
The addition of BAO data decreases the values of $\Omega_{m0}$ and $w_a$, and increases $w_0$, thereby intensifying the tension with the $\Lambda$CDM model.
The impact of data points at the redshift $z=0.51$ is small.
Including the spatial curvature as a free parameter, the constraints on the parameters $\Omega_{m0}$, $w_0$ and $w_a$ broaden a little compared to those in the flat case.
With the combined BAO+P18+H+D5 data,
we get $w_0=-0.797\pm 0.057$ and $w_a=-0.70^{+0.27}_{-0.23}$ for the flat CPL model,
$w_0=-0.786\pm 0.066$ and $w_a=-0.76^{+0.34}_{-0.28}$ for the non-flat flat CPL model,
the values of $\Delta$AIC reach $-10$ and $-8$ for the flat and non-flat cases, respectively,
indicating strong evidence for dynamical dark energy.
With the combined BAO+P18+H+U3 data,
we get $w_0=-0.775\pm 0.078$ and $w_a=-0.74^{+0.31}_{-0.25}$ for the flat CPL model,
$w_0=-0.754\pm 0.092$ and $w_a=-0.83^{+0.39}_{-0.34}$ for the non-flat flat CPL model,
the values of $\Delta$AIC are $-4$ and $-2$ for the flat and non-flat cases, respectively.
The results with the combined P18+H+PP data are more consistent with the $\Lambda$CDM model compared with those obtained from the combinations with D5 and U3.

For the SSLCPL model,
the constraints on $w_0$ from the combination with D5 and U3 are similar, although the error bars on $w_0$ from the combination with U3 are bigger that those from the combination with D5.
The constraints on $w_0$ obtained from the combined data with PP are consistent with the $\Lambda$CDM model at around the $1\sigma$ level,
while the constraints on $w_0$ from the combined data with D5 and U3 are consistent with the $\Lambda$CDM model at around the $2\sigma$ level.

In conclusion, the influence of DESI BAO data on the constraint of $w_0$ is small in the SSLCPL model.
Compared with the CPL model, the tension with the $\Lambda$CDM model is reduced a little bit for the SSLCPL model,
indicating that the evidence for dynamical dark energy from DESI BAO data is contingent on cosmological models.

\begin{acknowledgments}
This research is supported in part by the National Natural Science Foundation of China under Grant No. 12175184, the National Key Research and Development Program of China under Grant No. 2020YFC2201504 and the Chongqing Natural Science Foundation under Grant No. CSTB2022NSCQ-MSX1324.
\end{acknowledgments}

%

\end{document}